# Parallel Clustering of High-Dimensional Social Media Data Streams

Xiaoming Gao
School of Informatics and Computing
Indiana University
Bloomington, IN, USA
gao4@umail.iu.edu

Emilio Ferrara
School of Informatics and Computing
Indiana University
Bloomington, IN, USA
ferrarae@indiana.edu

Judy Qiu
School of Informatics and Computing
Indiana University
Bloomington, IN, USA
xqiu@umail.iu.edu

*Abstract*—We introduce Cloud DIKW as an analysis environment supporting scientific discovery through integrated parallel batch and streaming processing, and apply it to one representative domain application: social media data stream clustering. In this context, recent work demonstrated that high-quality clusters can be generated by representing the data points using high-dimensional vectors that reflect textual content and social network information. However, due to the high cost of similarity computation, sequential implementations of even single-pass algorithms cannot keep up with the speed of real-world streams. This paper presents our efforts in meeting the constraints of real-time social media stream clustering through parallelization in Cloud DIKW. Specifically, we focus on two system-level issues. Firstly, most stream processing engines such as Apache Storm organize distributed workers in the form of a directed acyclic graph (DAG), which makes it difficult to dynamically synchronize the state of parallel clustering workers. We tackle this challenge by creating a separate synchronization channel using a pub-sub messaging system (ActiveMQ in our case). Secondly, due to the sparsity of the high-dimensional vectors, the size of centroids grows quickly as new data points are assigned to the clusters. As a result, traditional synchronization that directly broadcasts cluster centroids becomes too expensive and limits the scalability of the parallel algorithm. We address this problem by communicating only dynamic changes of the clusters rather than the whole centroid vectors. Our algorithm under Cloud DIKW can process the Twitter 10% data stream ("gardenhose") in real-time with 96-way parallelism. By natural improvements to Cloud DIKW, including advanced collective communication techniques developed in our Harp project, we will be able to process the full Twitter data stream in real-time with 1000-way parallelism. Our use of powerful general software subsystems will enable many other applications that need integration of streaming and batch data analytics.

## I. Introduction

As data intensive computing problems evolve, many applications require integrated batch analysis and streaming analysis in the cloud. A good demonstration of the system-level innovation for supporting such use cases is Google's Cloud DataFlow [19]. Moving forward with this trend, we propose the Cloud DIKW (Data, Information, Knowledge, Wisdom) environment as shown in Fig. 1. It is designed to support analytic pipelines that require the integration of both sophisticated batch data processing algorithms and non-trivial streaming algorithms. By "non-trivial" algorithms, we refer to the cases where parallel workers not only process stream partitions independently, but also dynamically synchronize with the global state from time to time. The synchronization strategy could either leverage a pub-sub messaging system, or reuse the communication mechanisms in batch algorithms, or use a combination of both.

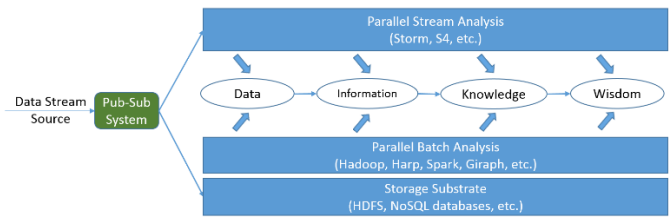

Fig. 1. Architecture of Cloud DIKW

This paper presents our work in applying this environment to support one representative application: clustering of social media data streams. As an important data mining technique, clustering is used in many applications involving social media stream analysis, such as meme [14][29], event [10], and social bots detection [14]. As an example, Fig. 2 illustrates the analysis pipeline of the DESPIC (Detecting Early Signatures of Persuasion in Information Cascades) platform [14] that is being developed by the Center for Complex Networks and Systems Research at Indiana University. This platform first clusters posts collected from social streams (e.g., tweets from Twitter) into groups of homogenous memes, according to various measures of similarity, and then uses classification methods to detect memes generated by real users and separate them from those produced by social bots [15].

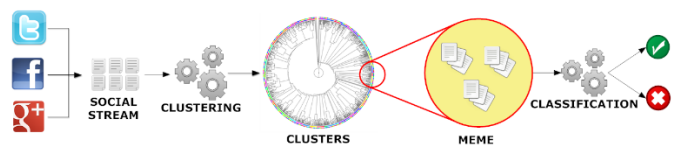

Fig. 2. DESPIC analysis pipeline for meme clustering and classification [14]

Social media data streams come in the form of continuous sequences of atomic posts, e.g. Twitter's tweets or Facebook's status updates. The target of the clustering process is to group messages that carry similar meaning together, while capturing the dynamic evolution of the streams that is closely related to social activities in the real world. For example, two tweets, "Step up time Ram Nation. #rowdyrams" and "Lovin @SpikeLee supporting the VCU Rams!! #havoc", should be grouped into the same cluster because they both talk about the VCU (Virginia Commonwealth University) basketball team. Furthermore, the appearance of "@SpikeLee" in the cluster is an indicator of the event that the famous director Spike Lee was wearing a VCU T-shirt while watching the VCU and UMass game courtside on Mar 16th, 2013.

In order to design a high-quality clustering algorithm, some unique characteristics of social posts must be considered. For instance, the length of the textual content of a social message is normally short, which makes clustering methods solely based on lexical analysis ineffective [10][20][29]. Social messages also carry rich information about the underlying social network (e.g. through the functionality of 'retweet' and 'mention' on Twitter), which can be valuable for measuring the similarity among data points and clusters. In addition they may contain other metadata such as temporal and geographical information, hashtags, URLs, etc., which can also be leveraged to effectively guide the clustering process. Fig. 3 illustrates an example tweet received from the Twitter Streaming API [35]. Besides textual content, hashtags, and URLs, each tweet also contains information about the creation time and geolocation of the source (i.e., the author of the tweet), the user(s) mentioned in the tweet, and possible retweet relationship between tweets.

```
{
    "text":"Lovin @SpikeLee supporting the VCU Rams!!  #HAVOC",
    "created_at":"Sat Mar 16 20:40:13 +0000 2013",
    "id_str":"313026943249444864",
    "entities":{
        "user_mentions":[{
            "screen_name":"SpikeLee",
            "id_str":"254218516",
            "name":"Spike Lee",
            ...}],
        "hashtags":[{
            "text": "HAVOC",
            ...}],
        "urls":[]},
    "user":{
        "created_at":"Sat Jan 22 18:39:46 +0000 2011",
        "friends_count":63,
        "id_str":"241622902",
        ...},
    "retweeted_status":null,
    "geo": {
        "type": "Point",
        "coordinates": [37.64760441, -77.60201846]},
    ...
}
```

Fig. 3. An example social message from Twitter Streaming API

Domain researchers in the area of social media data analysis have recently invested a great deal of efforts toward developing proper data representations and similarity metrics to generate high-quality clusters [10][14][20][29]. An important conclusion is that the data representation should not only describe the textual features of the social messages, but also capture the temporal, geographical, and social network information therein attached. For example, Aggarwal and Subbian [10] proposed an event-detection system that uses two high-dimensional vectors to describe each social post: one content vector that represents the textual word frequencies, and another binary vector housing the IDs of the social message's recipients (e.g., the followers of a tweet's author on Twitter). To compute the similarity between two social messages, an independent score is first computed using each vector, and then a linear combination of the two scores is taken as the overall similarity between the two messages. It has been demonstrated that the quality of the resulting clusters can be significantly improved by using the combined similarity rather than just the textual content similarity. JafariAsbagh et al. [29] proposed to first group the social messages into 'protomemes' according to shared metadata such as hashtags and URLs, and then use the protomemes as input data points to the clustering algorithm. They use four high-dimensional vectors to describe each protomeme, and define a new 'diffusion network' vector to replace the full followers vector used in [10], which is not easily available in a practical streaming scenario. The authors show that a combination of these new techniques can help generate better clustering results than previous methods when measured against a common ground truth data set.

To achieve efficient processing of social media data streams, these special data representations and similarity metrics are normally applied in a single-pass clustering algorithm such as online K-Means and its variants [2][10][29]. The algorithm can be further equipped with mechanisms like sliding time windows [1][29], weighted data points [2][8][9][11], and outlier detection [8][10][17][29] to deal with the dynamic evolution of the streams. However, due to the high cost of similarity computation coming from the high-dimensional vectors, sequential implementations of such single-pass streaming algorithms are not fast enough to match the speed of real-world streams. For example, the fastest implementation presented in [10] can only process less than 20,000 tweets per hour, while the Twitter gardenhose stream [25] generates over 1,000,000 tweets in one hour. According to a test we carried out, it takes 43.4 hours for a sequential implementation of the algorithm in [29] to process one hour's worth of data collected through the gardenhose Twitter streaming API. It becomes clear that parallelization is a necessity in order to handle real-time data streams.

In this paper we describe our work in parallelizing a state-of-the-art social media data stream clustering algorithm presented in [29], which is a variant of online K-Means incorporating a sliding time window and outlier detection mechanisms. We use Apache Storm [6] stream processing engine in Cloud DIKW for data transmission and work load distribution, and identify and tackle two system-level challenges emerging from parallelization of such algorithms.

The first challenge concerns the fact that most stream processing engines organize the distributed processing workers in the form of a directed acyclic graph (DAG); this makes it difficult to dynamically synchronize the state of the parallel clustering workers without breaking the "live" processing of the stream. The reason is that the synchronization step requires parallel workers to send their local updates either to each other or to a global updates collector which will then broadcast the updated global state back to the parallel workers. Both methods inevitably create cycles in the communication channel, which is not supported in the DAG-oriented stream processing frameworks. To address this challenge, we create a separate synchronization channel by incorporating the pub-sub messaging system ActiveMQ [5] into Cloud DIKW, and combine its functionality with Storm to coordinate the synchronization process.

The second issue is that the sparsity of high-dimensional vectors may cause the cluster centroids to greatly increase in size with the addition of new data points to the clusters. Fig. 4 illustrates a cluster containing two tweets about VCU basketball as mentioned earlier. Due to the sparsity of the content vector (assuming the hashtags and user mentions are extracted as another separate vector) of each data point, they only overlap along one dimension "ram". As a result, the length of the content vector of the centroid, which is computed as an average of the two data points, is close to the sum total length for two separate vectors. Due to the high dimensionality of these vectors, this trend can continue as more data points are added, and the length of the centroid vectors increases dramatically. A sliding time window mechanism may help to limit the total size by removing old data points, but the full centroids data remains large and difficult to transfer over the network. Consequently, the classic K-Means synchronization strategy of directly broadcasting the cluster centroids becomes infeasible and hampers scalability of the parallel algorithm. To solve this problem, we propose a new strategy that broadcasts the dynamic changes (i.e. the "deltas") of the clusters rather than the complete centroids data. Since the size of the deltas are small, we are able to keep the synchronization cost at a low level and achieve good scalability. For sake of simplicity, we name the traditional synchronization strategy the *full-centroids strategy*, and our new synchronization strategy the *cluster-delta strategy*.

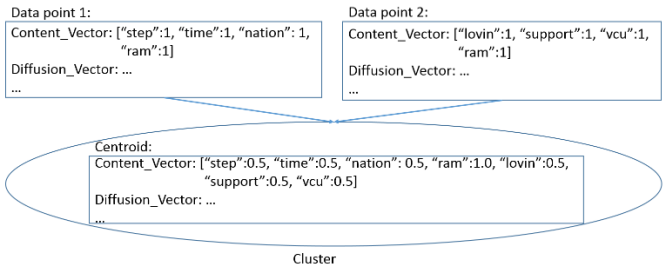

Fig. 4. An example of growing vector size of centroids

We use a real dataset collected through the Twitter streaming API 10% sample ("gardenhose") [25] to verify the effectiveness of our solutions and evaluate the scalability of our parallel algorithm. The results demonstrate that we can keep up with the speed of the Twitter gardenhose stream using less than 100 parallel clustering workers.

The rest of this paper is organized as follows. Section II discusses related work and their connections to our research. Section III gives a brief description of the sequential algorithm that we parallelize. Section IV explains the implementation of our parallel algorithm. Section V evaluates the effectiveness of our synchronization strategy and the scalability of our parallel algorithm. Section VI concludes with future work.

## II. RELATED WORK

Data stream clustering algorithms have been an active research area for many years as witnessed by Ding et al. review work [31]. For the problem of high-dimensional data stream clustering, techniques such as projected/subspace clustering [8][9][38] and density-based approaches [1][17][38] have been proposed and investigated. Due to the unique data representations (multiple high-dimensional vectors from totally independent spaces) and similarity metrics used for social media data streams, it seems hard to apply these existing techniques to the case of social media streams. We listed and discussed practical limitations in a previous work [14]. Here we inherit the high-dimensional data representation and similarity metrics that have been proven effective, and focus on improving the efficiency of the clustering algorithm through parallelization.

The algorithm presented in [10] uses sketch tables [12] to deal with the growing size of tweet followers network information maintained for the clusters. However, sketch tables only approximate vector values and thus may impact the accuracy of the clustering results. In the case of our algorithm, since the size of the centroid vectors is constrained by the size of the sliding time window, we are not forced to use sketch tables with their loss of accuracy. For faster data streams or longer time windows, a sketch table-based implementation could eventually become more efficient in terms of both space and time for computing the similarities between data points and cluster centroids. Nonetheless, our cluster-delta synchronization strategy may still achieve better efficiency than broadcasting the whole sketch tables in such cases since the sketch tables have to be large enough to ensure accuracy.

A similar work to ours is the parallel implementation of the Sequential Leader Clustering [22] algorithm presented in [18], which also leverages Storm [6] for parallel processing and data stream distribution. The parallel clustering algorithm by Wu et al. is simplified, because it only considers the textual content of social messages and uses Locality-Sensitive Hashing [4] to guide the stream distribution, which avoids synchronization among the parallel clustering workers. However, this type of algorithms is unable to make use of the valuable social network information contained in the data streams. Callau-Zori proposed a distributed data stream clustering protocol based on sequential (a, b)-approximation algorithms for the K-Means problem [27]. Although the author provides a theoretical analysis of its accuracy and efficiency, it does not address the special case of high-dimensional data, and only considers the situation within a single time window.

Compared with streaming databases such as Aurora [28] and Borealis [13], the functionality of our clustering workers in Storm is more complicated than their streaming operators for evaluating SQL queries. Cloud DIKW can utilize other stream processing engines such as Apache S4 [26] and Spark Streaming [30]. We choose Storm because its pull-based data transmission mode makes it easy to carry out controlled experiments at different levels of parallelism. Storm gives us more flexibility to implement and test different synchronization strategies. Interested readers may refer to [32] for a survey of major distributed stream processing frameworks.

## III. SEQUENTIAL CLUSTERING ALGORITHM

The sequential algorithm we parallelize was originally proposed in [29] for clustering memes in the Twitter streams of tweets. In order to generate high-quality clusters, the algorithm first groups tweets into 'protomemes', and then uses these protomemes as input data points for the clustering process. We start by introducing the definition of a protomeme and its data representation.

### A. Protomemes and Clusters

A *protomeme* is defined as a set of tweets grouped together according to a shared entity of one of the following types:

- **Hashtags**. Tweets containing the same hashtag.

- **Mentions**. Tweets mentioning the same user. A mention is identified by a user's screen name preceded by the '@' symbol in the text body of a tweet.

- **URLs**. Tweets containing the same URL.

- **Phrases**. Tweets sharing the same phrase. A phrase is defined as the textual content of a tweet that remains after removing the hashtags, mentions, URLs, and after stopping and stemming [23].

We call these four types of entities *markers* of protomemes. Note that according to this definition, a tweet may belong to multiple protomemes. Each protomeme is represented by its marker and four high-dimensional vectors:

(1) A binary *tid vector* containing the IDs of all the tweets in this protomeme: $V_T = [tid_1 : 1, tid_2 : 1, …, tid_T : 1]$;

(2) A binary *uid vector* containing the IDs of all the users who authored the tweets in this protomeme: $V_U = [uid_1 : 1, uid_2 : 1, …, uid_U : 1]$;

(3) A *content vector* containing the combined textual word frequencies for all the tweets in this protomeme: $V_C = [w_1 : f_1, w_2 : f_2, …, w_C : f_C]$;

(4) A binary vector containing the IDs of all the users in the *diffusion network* of this protomeme. The diffusion network of a protomeme is defined as the **union** of the set of tweet authors, the set of users mentioned by the tweets, and the set of users who have retweeted the tweets. We denote this *diffusion vector* as $V_D = [uid_1 : 1, uid_2 : 1, …, uid_D : 1]$.

A *cluster* is defined as a set of protomemes grouped together according to a certain similarity metric. Since a tweet may belong to multiple protomemes, clusters may overlap in terms of tweets. The centroid of each cluster is represented with four high-dimensional vectors, which are the averages of corresponding vectors of all protomemes in the cluster. We denote the vectors of the cluster centroid as $V_T$, $V_U$, $V_C$, and $V_D$.

To compute the *similarity* between a protomeme and a cluster, the **cosine similarity** between each vector of the protomeme and the corresponding vector of the cluster centroid is first computed. Then the **maximum value** of all these cosine similarities is chosen as the overall similarity between the two. We demonstrated in [14] that, for the purpose of generating high-quality clusters, taking the maximum similarity score is as effective as using an optimal linear combination of all the similarity scores. There are multiple ways to define *distance* based on the similarity; we use the simplest form $1 - similarity$.

### B. Sequential Clustering Algorithm

---
**Algorithm** *TweetStreamClustering*
**Input parameters**:
  *K*: number of clusters;
  *t*: length of time step by which the time window advances;
  *l*: length of the time window in steps;
  *n*: number of standard deviations from the mean to identify outliers;
**begin**
 Initialize global list of clusters *cl* as empty;
 Initialize global list of protomemes *gpl* as empty;
 Initialize the time window *tw* as empty;
 Initialize $\mu$, $\sigma$ to 0;
 **while not** end of stream **do**
  advance the time window *tw* by *t*;
  **let** *npl* = list of protomemes generated from the tweets in *t*;
  **if** *cl* is empty **then**
   initialize *cl* using *K* random protomemes in *npl*;
   remove these *K* protomemes from *npl*;
  **endif**
  **for** each protomeme *p* in *gpl* that is older than the current *tw*
   delete *p* from *gpl* and the cluster it belongs to;
  **endfor**
  **for** each new protomeme *p* in *npl*
   **if** *p.marker* has been previously assigned to a cluster *c* **then**
    add *p* to *c* and update the centroid of *c*;
   **else**
    **let** *c* = the cluster in *cl* whose centroid is most similar to *p*;
    **if** $sim(p, c) > \mu - n * \sigma$ **then**
     add *p* to *c* and update the centroid of *c*;
    **else**
     create a new cluster *c'* containing only one protomeme *p*;
     **if** there is an empty cluster in *cl* **then**
      replace the empty cluster with *c'*;
     **else**
      replace the least recently updated cluster in *cl* with *c'*;
     **endif**
    **endif**
   **endif**
   add *p* to *gpl*;
   dynamically maintain $\mu$ and $\sigma$;
  **endfor**
 **endwhile**
**end**

---

Fig. 5. The social media stream clustering algorithm from [29]

Fig. 5 illustrates the sketch of the sequential clustering algorithm from [29]. The algorithm controls its progress through a sliding time window that moves step by step. The length of a time step in seconds and the length of the time window in steps are given as input parameters. These are defined with respect to the timestamps of the social posts (i.e., the tweets), not the wall-clock time for running the algorithm. Every time the sliding window advances, old protomemes falling out of the current window are deleted from the clusters and new ones are generated using the tweets from the latest time step. For every new protomeme, the algorithm first checks whether others with the same marker have been previously assigned to a cluster. If so, the new protomeme will be added to the same cluster. Otherwise, the algorithm will compute the new protomeme's similarity with all the existing clusters, and decide whether or not this is an outlier. If not, the protomeme is assigned to the most similar cluster. Otherwise, a new cluster is created and initialized with this new protomeme, then inserted into the list of all clusters by replacing either an empty cluster or the least recently updated one. In order to determine whether the protomeme is an outlier, the algorithm maintains the mean $\mu$ and standard deviation $\sigma$ of the similarities between all processed protomemes and the centroid of the clusters they belong to. If the similarity between a new protomeme and its closest cluster is smaller than the mean by more than $n$ standard deviations, then the protomeme is identified as an outlier. $\mu$ and $\sigma$ are maintained incrementally as in [10].

The quality of clusters generated by this algorithm was evaluated in [29] using a ground truth dataset collected from the Twitter gardenhose stream [25] during a week in 2013, which includes all the tweets containing the Twitter trending hashtags [36][16] identified during that time. A variant of the *Normalized Mutual Information* (NMI) [24] measurement, LFK-NMI [3], which is especially well suited for the case of overlapping clusters, was computed between the result clusters of the algorithm and the ground truth clusters. The results in [29] show that this algorithm can achieve better performance than previous state-of-the-art methods, including the one presented in [10]. We use the same ground truth dataset and LFK-NMI measurement to verify the effectiveness of our parallel implementation of the algorithm in Section V.

### C. Opportunities and Difficulties for Parallelization

We run the sequential algorithm on a raw dataset (without any filtering) containing six minutes of tweets (2014-08-29 05:00:00 to 05:05:59) collected from the Twitter gardenhose stream. By fixing the parameters $K$, $l$, and $n$ to 120, 6, and 2, and varying the length of a time step, we collect some important runtime statistics that are informative to the development of the parallel version of the algorithm.

Table I presents the results for the last time step of the whole clustering process when the time step length is increased from 10 to 30 seconds (which means the time window length is increased from 60 to 180 seconds). The numbers for the other time steps follow a similar pattern. The second column measures the total length of the content vectors of all the cluster centroids at the end of the last time step; the third column measures the time spent on computing the similarities between protomemes and cluster centroids in that time step; and the fourth column measures the time spent on updating the vectors of the cluster centroids.

TABLE I. RUNTIME STATISTICS FOR THE SEQUENTIAL ALGORITHM

| Time Step Length (s) | Total Length of Content Vector | Similarity Compute time (s) | Centroids Update Time (s) |
|---|---|---|---|
| 10 | 47749 | 33.305 | 0.068 |
| 20 | 76146 | 78.778 | 0.113 |
| 30 | 128521 | 209.013 | 0.213 |

Some interesting observations lead to our research of parallelizing the streaming algorithm: first, the whole clustering process is dominated by the computation of similarities. The ratio of **similarity compute time / centroids update time** in Table I increases from 490 to 981 as the length of the time window increases from 10 to 30 secs. This implies the feasibility of parallelizing the similarity computation, and processing the global updates of centroids with a central collector. Furthermore, the longer the time window, the more we can benefit from parallelization.

We also observed that the content vector size of the centroids expands as the length of the time window increases. In fact, the other vectors ($V_T$, $V_U$, $V_D$) demonstrate the same trend. This confirms our analysis in Section I about the infeasibility of traditional synchronization strategies. To address this issue, we design the new cluster-delta strategy, which will be presented in Section IV.

### IV. PARALLEL IMPLEMENTATION ON STORM

#### A. Storm

Apache Storm is a stream processing engine designed to support large-scale distributed processing of data streams. It defines a stream as an unbounded sequence of *tuples*, and provides an easy-to-use event-driven programming model to upper level applications. Stream processing applications are expressed as *topologies* in Storm. There are two types of *processing elements* in a topology, *spouts* and *bolts*, which are organized into a DAG through the streams connecting them. A spout is a source of streams that generates new tuples and injects them into the topology. A bolt can consume any number of input streams, do some processing to each tuple of the streams, and potentially generate and emit new tuples to the output streams. To define a topology, an application only needs to provide implementation logic for spouts and bolts, specify the runtime parallelism level of each type, and configure the data distribution patterns among them. The Storm framework will automatically take care of system management issues including data transmission, parallel spouts/bolts execution, work load distribution, and fault tolerance.

Fig. 6 illustrates the standard architecture of a Storm cluster. The whole cluster consists of two types of nodes: one master node and multiple worker nodes. The master node runs a daemon process called *Nimbus* responsible for assigning spout and bolt tasks to the worker nodes and monitoring their status for failures. Every worker node runs a *Supervisor* daemon process, which manages the resources on the local node, and accepts task assignments from the *Nimbus*. Spout and bolt tasks are launched as parallel threads in *worker processes*. The number of worker processes on each node is configurable as a system parameter. The number of threads to run for each type of spout and bolt in a topology can be configured through the parallelism parameters. Fault tolerant coordination between the *Nimbus* and the *Supervisors* uses Zookeepers [7].

Storm adopts the 'pull-based' message passing model between the processing elements. Bolts pull messages from the upstream bolts or spouts. This ensures that bolts will never get excessive workload that they cannot handle. Therefore, overflow can only happen at the spouts. This model allows us to test our algorithm easily at different levels of parallelism. For example, we can implement spouts that generate streams by reading data from a file, and control their paces based on the number of acknowledgements received for tuples that have been processed. This will prevent the topology from getting overwhelmed by too much data no matter how slowly the bolts are working.

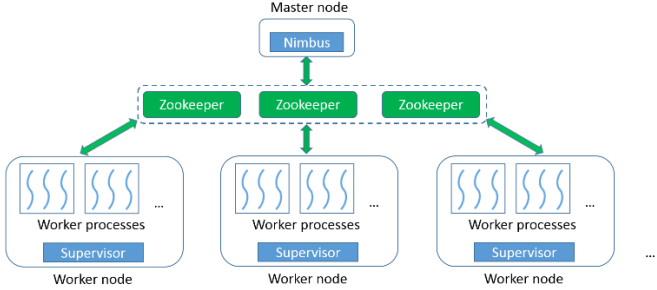

Fig. 6.  Storm architecture

### B. Implementation with Cluster-Delta Synchronization

Cloud DIKW implements the parallel version of the algorithm in a Storm topology, as illustrated in Fig. 7. There is one type of spout, *Protomeme Generator Spout*, and two types of bolts, *Clustering Bolt* and *Synchronization Coordinator Bolt*. For simplicity, we call them *protomeme generator*, *cbolt*, and *sync coordinator*. At runtime, there is one instance of the protomeme generator, multiple instances of cbolts working in parallel, and one instance of sync coordinator. A separate synchronization channel is created between the cbolts and the sync coordinator using the ActiveMQ pub-sub messaging system [5]. ActiveMQ allows client applications to connect to *message brokers*, and register themselves as *publishers* or *subscribers* to various *topics*. Publishers can produce messages and publish them to a certain topic, and the message broker will automatically deliver the messages to all the subscribers of that topic. In our topology, the sync coordinator is registered as a publisher to a topic named "clusters.info.sync", and all the cbolts are registered as subscribers to this topic. The lifetime of the whole topology can be divided into two phases, an *initialization phase* and a *running phase*. We introduce the working mechanism of each type of spout and bolt in both phases.

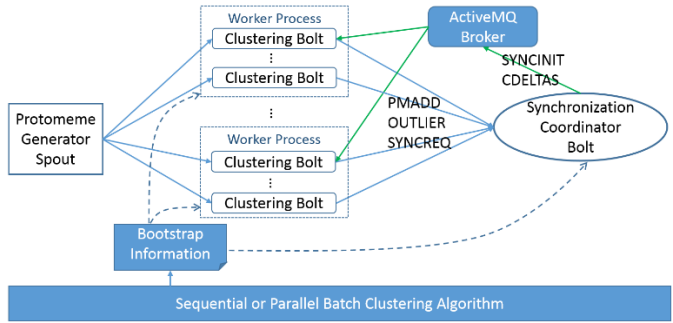

Fig. 7.  Storm topology for the parallel stream clustering algorithm

### Protomeme Generation and Distribution

During the **initialization phase**, every processing element reads some information from a bootstrap file. The protomeme generator reads the start time of the current time step, the length of a time step in seconds, and the length of a time window in steps. After reading this information, the generator can either connect to an external stream of tweets or open a file containing tweets for generating protomemes.

During the **running phase**, the protomeme generator keeps reading and buffering tweets for the "current" time step, until it identifies a tweet falling into the next time step. Then it generates protomemes using the buffered tweets. Every protomeme is associated with a *creation timestamp* and an *ending timestamp*, which are set based on the timestamp of the earliest and latest tweet in the protomeme. To facilitate the construction of diffusion vectors of protomemes, an **in-memory index structure** is maintained to record the mapping between each tweet ID and the set of user IDs who have retweeted it. To construct the diffusion vector of a protomeme, the user IDs of the tweet authors and the user IDs mentioned in its tweets are first added to the vector. Then the index is queried for each tweet ID of the protomeme, and the corresponding user IDs found in the index are added to the vector. The protomeme generator emits one tuple to its output stream for every newly generated protomeme. The tuples are evenly distributed among all the parallel cbolts based on the hash values of their markers. Therefore, protomemes generated in different time steps but sharing the same marker will always be processed by the same cbolt.

### Protomeme Clustering

During the **initialization phase**, the cbolts and sync coordinator first read the same time window parameters as the protomeme generator; then they read the input parameter *n* (number of standard deviations for outlier detection), and a list of initial clusters. The initial clusters are generated by running either a parallel batch clustering algorithm, or the sequential stream clustering algorithm over a small batch of data from recent history. The initial values of $\mu$ and $\sigma$ are then generated based on the protomemes contained in the initial clusters.

During the **running phase**, protomemes are processed in small batches. A *batch* is defined as the number of protomemes to process, which is normally configured to be much smaller than the total number of protomemes in a single time step. Upon receiving a protomeme, the cbolt first checks its creation timestamp to see if it starts a new time step. If so, the cbolt will first advance the current time window by one step, and delete all the old protomemes falling out of the time window from the clusters. Then it performs the outlier detection procedure and protomeme-cluster assignment in the same way as in the sequential algorithm, based on the current clusters and $\mu$, $\sigma$ values. If the protomeme is an outlier, an *OUTLIER* tuple containing the protomeme will be emitted to the sync coordinator. If it can be assigned to a cluster, a *PMADD* tuple will be emitted. Note that the cbolt does not immediately create a new cluster with the outlier protomeme, because outlier protomemes detected by different cbolts may be similar to each other and thus should be grouped into the same cluster. Such global grouping can only be done by the sync coordinator, which collects *OUTLIER* tuples generated by all the cbolts. For the case of *PMADD*, the centroid of the corresponding cluster is not immediately updated either. Instead, clusters are only updated during the synchronization between two consecutive batches. This ensures that within the same batch, different cbolts are always comparing their received protomemes against the same set of global clusters.

Within each batch, the sync coordinator maintains a list of "cluster delta" data structures and another list of outlier clusters. Upon receiving a *PMADD*, it will simply add the protomeme contained in the tuple to the delta structure of the corresponding cluster, and change the latest update time of the delta structure to the ending timestamp of the protomeme in case the ending timestamp is larger. Since the sync coordinator collects *PMADD* from all parallel cbolts, the delta structures will contain the global updates to each cluster. For an *OUTLIER* tuple, it will first check whether the protomeme contained in the tuple can be assigned to any existing outlier cluster. If so, it is simply added to that outlier cluster; otherwise a new outlier cluster is created and appended to the list of outlier clusters. After processing each tuple, the values of $\mu$ and $\sigma$ are dynamically updated.

*Synchronization*

As a final step of the **initialization phase**, the cbolts and sync coordinator connect to an ActiveMQ message broker and register as subscribers and the publisher. Since the cbolt tasks run as threads in worker processes, they first go through an **election** step to select one **representative thread** within each process. Only the representative thread will be registered as a subscriber, and the synchronization message received will be shared among the threads in the same process. This election step can significantly reduce the amount of data transmission caused by synchronization.

At the **running phase**, a synchronization procedure is launched when the number of processed protomemes reaches the batch size. The whole procedure consists of three steps as detailed in Fig. 8: *SYNCINIT*, *SYNCREQ*, and *CDELTAS*. The *SYNCINIT* step initiates the procedure and notifies the cbolts to start synchronization. In the *SYNCREQ* step, each cbolt will temporarily stop processing incoming protomemes, and emit a *SYNCREQ* tuple. After receiving *SYNCREQ* from all the cbolts, the sync coordinator will sort the deltas of all the clusters (including the outlier clusters) by the latest update time, and pick the top *K* with the highest values to construct a *CDELTAS* message, which also contains latest global values of $\mu$ and $\sigma$. The message is then published through ActiveMQ. Upon receiving *CDELTAS*, every cbolt will update their local copy of clusters and $\mu$, $\sigma$ values to a new global state, then resume processing the protomemes for the next batch. Note that the *SYNCINIT* step and the temporary stopping of the cbolts are necessary to ensure that protomemes processed by different cbolts and received by the sync coordinator are always handled with regards to the same global view of the clusters. Since the size of *CDELTAS* is normally small and stable, the synchronization step can usually finish in a short time, as will be demonstrated in Section V.

In order to achieve the best performance for the whole synchronization procedure, an optimal solution for *SYNCINIT* is also necessary. We tested three methods in this regard. With **spout initiation**, the protomeme generator counts the number of protomemes emitted and broadcasts a *SYNCINIT* tuple through Storm when the batch size is reached. With **cbolt initiation**, each cbolt counts the number of protomemes processed by itself and directly emits a *SYNCREQ* tuple when it reaches the expected average. This method is similar to the synchronization mechanism used in typical iterative batch algorithms. However, due to the buffering effect of Storm and varied processing speed among cbolts, both methods suffer from a large variance in the *SYNCREQ* time observed by different cbolts. The variance can reach the level of seconds and totally eliminate the benefits of the cluster-delta strategy. This suggests that, due to the dynamic nature of streaming analysis, synchronization should be handled differently than in batch algorithms. To address this issue, we propose **sync coordinator initiation** as illustrated in Fig. 8. In this method, the sync coordinator counts the total number of *PMADD* and *OUTLIER* received, and publishes a *SYNCINIT* message using ActiveMQ if the batch size is reached. Because of the pushing-mode of message delivery and the small size of the message, it can be received by the cbolts within milliseconds. Therefore the large variance problem is avoided.

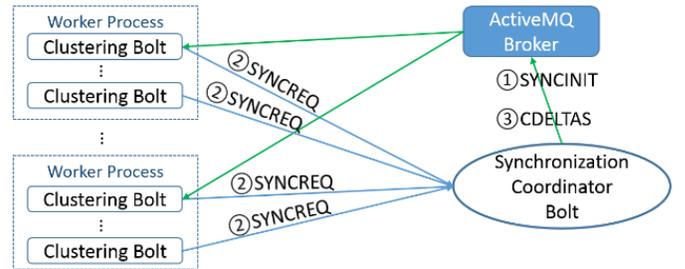

Fig. 8. Storm topology for the parallel stream clustering algorithm

*C. Implementation with Full-Centroids Synchronization*

To verify the effectiveness of our cluster-delta synchronization strategy, we implement another version of the parallel algorithm using the full-centroids strategy for comparison. The protomeme generation and processing logics of the full-centroids version are mostly the same as the cluster-delta version. There are, however, major differences in the implementation caused by the full-centroids strategy: during the processing time of each batch, the sync coordinator will maintain a full list of existing clusters, instead of their delta structures. During the synchronization time, instead of the *CDELTAS* message, it will generate a *CENTROIDS* message, which contains the whole centroid vectors of the clusters with the top *K* latest update times. Upon receiving the *CENTROIDS* message, every cbolt will use the centroid vectors contained in the message to replace the centroids of the old clusters.

Since the cbolt receives the centroid vectors rather than the incremental protomemes of each cluster, it can no longer maintain a full record of all the protomemes in the clusters. Therefore, the task of new time step detection and old protomeme deletion is moved to the sync coordinator. Since the centroids update time is negligible, if compared to the similarity compute time, this has little impact on the overall performance of the algorithm.

## V. EVALUATION OF THE PARALLEL ALGORITHM

We verify the correctness of our parallel algorithm by comparing its results with the sequential implementation, and evaluate its efficiency and scalability through comparison with the full-centroids synchronization strategy. Our tests run on a private eight-node cluster called "Madrid". The hardware configuration of each node is listed in Table II. Each node runs RHEL 6.5 and Java 1.7.0_45. The Cloud DIKW environment is constructed with Hadoop 2.5.0, HBase 0.96.0, Storm 0.9.2, and ActiveMQ 5.4.2. An ActiveMQ broker is deployed on the same node where the Storm Nimbus runs. Each node is configured to run at most four Storm worker processes, and the parallel instances of spouts and bolts are launched as threads spawned by these worker processes. The maximum heap size of each worker process is set to 11GB. Message compression with zip is enabled for ActiveMQ, and only one message broker is used in all tests of the parallel implementations.

TABLE II. PER-NODE HARDWARE CONFIGURATION OF MADRID

| CPU | RAM | Hard Disk | Network |
|---|---|---|---|
| 4 * 4 Quad-Core AMD Opteron 8356 2.3G Hz | 48GB | 4 TB HDD + 1TB SSD | 1Gb Ethernet |

### A. Correctness Verification

To test the correctness of our algorithm, we use the same ground truth dataset and LFK-NMI measurement as [29]. The LFK-NMI value is a number between 0 and 1 that indicates the degree of matching between two sets of result clusters. A value of 1 corresponds to a perfect matching, while a value of 0 means that the two sets of clusters are completely disjoint. The ground truth dataset was collected from the Twitter gardenhose stream [25] within the week of 2013-03-23 to 2013-03-29. It includes all the tweets containing the Twitter trending hashtags [36][16] identified during that time.

We first define the ground truth clusters as the sets of tweets corresponding to each trending hashtag: all tweets sharing a common trending hashtag are grouped into one separate cluster. Note that, since a tweet may contain multiple trending hashtags, the ground truth clusters may have overlaps. We then remove the trending hashtags from the content of all tweets, and run both the sequential implementation from [29] and our parallel implementation over the remaining dataset. As a result, protomemes corresponding to the trending hashtags will not be created and used as input data points to the clustering process. This is done to avoid giving an unfair advantage to protomeme-based algorithms that use hashtag information. Finally, we compute three LFK-NMI values: results of the sequential algorithm versus the ground truth clusters, results of the parallel algorithm versus the ground truth clusters, and results of the sequential versus the parallel algorithm. We use the same input parameters as the experiments completed in [29]: $K = 11$, $t = 60$ minutes, $l = 6$, and $n = 2$. For the parallel algorithm, we use two parallel cbolts and a batch size of 40.

Table III presents the LFK-NMI scores using the final clusters generated by the two algorithms. The high value of 0.728 in the first column indicates that the clusters generated by our parallel implementation match very well with the results of the original sequential implementation in [29]. Moreover, values in the second and third column suggest that, when measured against the same ground truth clusters, our parallel implementation can achieve a degree of matching comparable or better (we observe an improvement of around 10%) than the sequential implementation. These scores show that our parallel implementation is correct and can generate results that are consistent with the sequential algorithm. The value 0.169 is consistent with the original test results in [29]. Furthermore, the slightly higher value of 0.185 indicates that processing the protomemes in small batches may be helpful for improving the quality of the clusters.

TABLE III. LFK-NMI VALUES FOR CORRECTNESS VERIFICATION

| Parallel vs. Sequential | Sequential vs. ground truth | Parallel vs. ground truth |
|---|---|---|
| 0.728 | 0.169 | 0.185 |

### B. Performance Evaluation

To evaluate the performance and scalability of our parallel algorithm in Cloud DIKW, we use a raw dataset collected from the Twitter gardenhose stream without applying any type of filtering. It contains a total number of 1,284,935 tweets generated within one hour (from 05:00:00 AM to 05:59:59 AM) on 2014-08-29. We first run the sequential algorithm over the whole dataset using input parameters $K = 240$, $t = 30$ seconds, $l = 20$, and $n = 2$, and measure the total processing time. Note that the time window has a length of 10 minutes and thus may contain a large number of protomemes. Then we run the two parallel implementations at different levels of parallelism, and measure their processing time, speedup, and other important statistics. We use the clusters generated for the first 10 minutes of data as the bootstrap clusters, and process the following 50 minutes of data using the parallel algorithms. The average number of protomemes generated in each time step is 19908, and the batch size is set to 6144.

The total processing time of the sequential algorithm is 156,340 seconds (43.43 hours), and the time spent on processing the last 50 minutes of data is 139,950 seconds (38.87 hours). Fig. 9 compares the total processing time of the two parallel implementations, and some important statistics are given in Table IV and V. Numbers in brackets in the first column tell how many Storm worker processes were used for hosting the cbolt threads. These correspond to the total numbers of ActiveMQ receivers in each run. Here we list the numbers that delivered the best overall performance. The length of the synchronization message in the last column is measured before ActiveMQ runs any compression. Fig. 10 compares the scalability of the two parallel implementations (the blue line and the red line).

TABLE IV. STATISTICS FOR FULL-CENTROIDS VERSION

| Number of cbolts (worker processes) | Comp time / sync time | Sync time per batch (sec) | Avg. length of sync message |
|---|---|---|---|
| 3   (1) | 31.56 | 6.45 | 22113520 |
| 6   (1) | 15.53 | 6.51 | 21595499 |
| 12  (2) | 7.79 | 6.60 | 22066473 |

| Number of cbolts (worker processes) | Comp time / sync time | Sync time per batch (sec) | Avg. length of sync message |
|---|---|---|---|
| 24 (4) | 3.95 | 6.76 | 22319413 |
| 48 (7) | 1.92 | 7.09 | 21489950 |
| 96 (28) | 0.97 | 8.77 | 21536799 |

TABLE V.  STATISTICS FOR CLUSTER-DELTA VERSION

| Number of cbolts (worker processes) | Comp time / sync time | Sync time per batch (sec) | Avg. length of sync message |
|---|---|---|---|
| 3 (1) | 289.18 | 0.54 | 2525896 |
| 6 (1) | 124.62 | 0.56 | 2529779 |
| 12 (2) | 58.45 | 0.58 | 2532349 |
| 24 (4) | 27.44 | 0.64 | 2544095 |
| 48 (7) | 11.96 | 0.76 | 2559221 |
| 96 (28) | 5.95 | 0.89 | 2590857 |

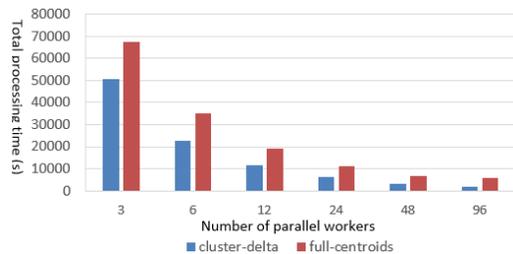

Fig. 9. Total processing time: cluster-delta vs. full-entroids

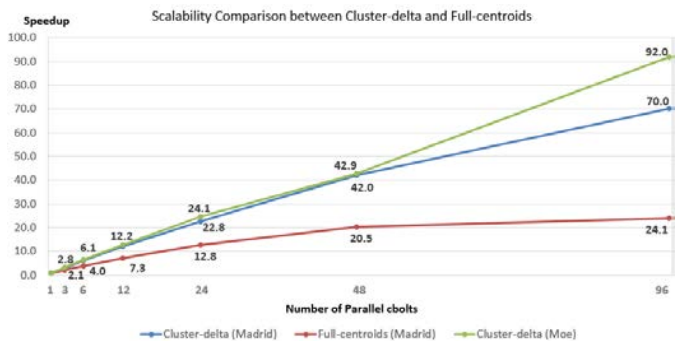

Fig. 10. Scalability comparison between cluster-delta and full-entroids

Table IV demonstrates that due to the large size of the cluster centroids, the full-centroids strategy generates a large synchronization message over 20 MB, and incurs a long synchronization time in every batch. In addition, the synchronization time increases as the number of parallel cbolts increases, because the single ActiveMQ broker needs to send a large message to more subscribers. The total processing time for the case of 96 parallel cbolts is dominated by synchronization. As a result, the full-centroid algorithm demonstrates poor scalability, and stops getting faster after 48 parallel cbolts.

In comparison, the cluster-delta strategy generates a much smaller synchronization message, and thus keeps the per-batch synchronization time at a low level, as shown in Table V. The zip compression of ActiveMQ provides a compression ratio of about 1:6, so the actual message size sent over the network is less than 500KB. As the number of parallel cbolts increases, the computation time covers the major part of the total processing time for all cases. The parallel implementation using the cluster-delta strategy can achieve a near-linear scalability for up to 48 parallel cbolts. Overall, it demonstrates sub-linear scalability. Using 96 parallel cbolts, it finishes processing the 50 minutes' worth of data in 1,999 seconds (33.3 minutes), thus keeping up with and surpassing the speed of the Twitter gardenhose stream. Note that even for the case of 96 parallel cbolts, the per-batch synchronization time is still relatively low. A major reason for the relatively low speedup of 70.0 is lack of computation, because each cbolt only processes about 64 protomemes per batch. In case of longer time steps or faster data rate, it is possible to extend the near-linear-scalability zone to larger numbers of parallel cbolts by increasing the batch size. To verify this, we use a dataset containing 2,258,821 tweets for 1h (1:00:00 PM to 2:00:00 PM) on 2014-08-29, and run the same tests on a different computer cluster called "Moe" with better CPU and network configuration (Table VI). 1-2pm is the peak hour of day when gardenhose generates the most tweets. The average number of protomemes in each time step is 35358, and we set the batch size to 12288. The speed-ups are illustrated by the green line in Fig. 10. Due to larger *CDELTAS* messages, the sync time per batch for 96 parallel cbolts increases to 0.979 seconds, despite the faster network. However, since the batch size is large, we are able to retain the near-linear scalability, and finish 50 minutes' worth of data in 2345 seconds (39 minutes).

TABLE VI.  PER-NODE HARDWARE CONFIGURATION OF MOE

| CPU | RAM | Hard Disk | Network |
|---|---|---|---|
| 5 * Intel 8-core E5-2660v2 2.20GHz | 128GB | 48 TB HDD + 120GB SSD | 10Gb Ethernet |

## VI. CONCLUSIONS AND FUTURE WORK

Cloud DIKW is an analysis environment that supports integrated batch and streaming processing. We have investigated it to efficiently support parallel social media stream clustering algorithms. This research leads to some important conclusions. Firstly, the distributed stream processing framework used in Cloud DIKW provides a convenient way to develop and deploy large-scale stream processing applications. Yet in order to properly coordinate the dynamic synchronization between parallel processing workers, their DAG-oriented processing models need additional coordination tools that we successfully implemented here with pub-sub messaging. Generalizing this to other applications is an important research area and could lead to such desirable synchronization capability being added to Apache Storm

Moreover, the parallelization and synchronization strategies may differ depending on the data representations and similarity metrics of the application. For example, we observed that the high-dimensionality and sparsity of the data representation in our application led to nontrivial issues addressed here for both computation and communication. By replacing the traditional full-centroids synchronization strategy with the new cluster-delta strategy, our parallel algorithm is able to be scalable, and keep up with the speed of the real-time Twitter gardenhose stream with less than 100 parallel workers.

There are several interesting directions that we will explore in the future. We will integrate advanced collective communication techniques as implemented by the Iterative MapReduce Hadoop plugin Harp [33] into Cloud DIKW, and use them to improve the synchronization performance of both batch and streaming algorithms. Instead of using a "gather and broadcast" communication model, Harp can organize the parallel workers in a communication chain, so that the local updates generated by each worker can be transmitted through all the other workers in a pipeline. According to our earlier attempts [37] to apply this technique in the Twister iterative MapReduce framework [21], it can significantly reduce the synchronization time and ensure that the algorithm achieves near linear scalability. With improved synchronization speed, we can process the data at the rate of the whole Twitter firehose stream [34], which is about 10 times larger than gardenhose. To support higher data speed and larger time window sizes, we may apply the sketch table technique as described in [12] in the clustering bolts and evaluate its impact on the accuracy and efficiency of the whole parallel program. Variations in arrival rate and jitter in event distribution exist in many real-time data streams. Therefore, we will also make the parallel algorithm elastic to accommodate this irregularity in event arrival.

ACKNOWLEDGMENT

We gratefully acknowledge support from National Science Foundation grant OCI-1149432 and DARPA grant W911NF-12-1-0037. We thank Mohsen JafariAsbagh and Onur Varol for their generous help in explaining their sequential algorithm, and professors Alessandro Flammini, Geoffrey Fox and Filippo Menczer for their support and advice throughout the project.